# On-chip TIRF nanoscopy by applying Haar wavelet kernel analysis on intensity fluctuations induced by chip illumination

**Nikhil Jayakumar[1,*], Øystein I. Helle[1], Krishna Agarwal[1], Balpreet Singh Ahluwalia[1,2,#]**

[1]*Department of Physics and Technology, UiT The Arctic University of Norway, Tromsø 9037, Norway*

[2]*Department of Clinical Science, Intervention and Technology, Karolinska Insitute, 17177 Stockholm, Sweden*

[*]nik.jay.hil@gmail.com  [#]balpreet.singh.ahluwalia@uit.no

**Abstract:** Photonic-chip based TIRF illumination has been used to demonstrate several on-chip optical nanoscopy methods. The sample is illuminated by the evanescent field generated by the electromagnetic wave modes guided inside the optical waveguide. In addition to the photokinetics of the fluorophores, the waveguide modes can be further exploited for introducing controlled intensity fluctuations for exploitation by techniques such as super-resolution optical fluctuation imaging (SOFI). However, the problem of non-uniform illumination pattern generated by the modes contribute to artifacts in the reconstructed image. To alleviate this problem, we propose to perform Haar wavelet kernel (HAWK) analysis on the original image stack prior to the application of (SOFI). HAWK produces a computational image stack with higher spatio-temporal sparsity than the original stack. In the case of multimoded non-uniform illumination patterns, HAWK processing bre aks the mode pattern while introducing spatio-temporal sparsity, thereby differentially affecting the non-uniformity of the illumination. Consequently, this assists nanoscopy methods such as SOFI to better support super-resolution, which is otherwise compromised due to spatial correlation of the mode patterns in the raw image. Furthermore, applying HAWK prior to SOFI alleviates the problem of artifacts due to non-uniform illumination without degrading temporal resolution. Our experimental results demonstrate resolution enhancement as well as reduction in artifacts through the combination of HAWK and SOFI.

## 1. Introduction

In far-field optical microscopy, the diffraction of light limits the ability of the system to image two adjacent point sources distinctly to ∽ 200 nm. This is referred to as the resolution limit of the optical system. The theoretical resolution limit of an optical microscope is approximately $\frac{\lambda}{2NA}$, where $\lambda$ is the wavelength of fluorescent emission and NA is the numerical aperture of the optical system. A lot of features and physiological processes of interest lie below this resolution limit and hence, medical solutions to real-world problems require a resolution beyond this barrier. Different imaging methodologies such as near-field scanning methods [1, 2] and electron microscope (EM) [3, 4] supports much better resolution. Near-field methods are challenging as it requires bringing the probe close to the target and EM are incompatible with live cell imaging applications. Therefore, the invention of fluorescence based far-field super-resolution optical microscopy, commonly referred to as optical nanoscopy has gained popularity during the last two decades. Fluorescence microscopy methods are live cell compatible and allows selective imaging of cellular components via molecule-specific labeling in both fixed and living samples. High specificity, live-sample compatibility and visualization of structures below the resolution limit make fluorescence based optical nanoscopy popular among biologists [5].

For applications where high contrast imaging close to the membrane surface is required with reduced photo-toxicity and excellent optical sectioning, total internal reflection fluorescence (TIRF) microscopy is employed [6]. Conventionally, a high numerical aperture (N.A.) and a high magnification TIRF lens is used in TIRF microscopy, which limits the field of view (FoV). Recently, it was demonstrated that photonic-chip based TIRF microscopy enables TIRF imaging over large FoV and supports scalable resolution and FoV [7]. Furthermore, it was demonstrated that waveguide platforms fabricated using high refractive index contrast (HIC) materials are attractive as it can generate high intensity in the evanescent field [8]. The core of the optical waveguide is made of a high-refractive index material, with a top and a bottom cladding layer of lower refractive index material, and guides visible light due to total-internal reflection (TIR) at the core-cladding interface. The TIR of light at the core-cladding interface is accompanied with the generation of an evanescent field that exponentially decays in the cladding region. The limited penetration depth of the evanescent field from the core-cladding interface helps prevent out-of-focus light when harnessed for fluorescence excitation of a specimen lying on the chip surface. As a result, several imaging methods have been implemented using waveguide platforms over the past few years – direct stochastic optical reconstruction microscopy (*d*STORM) [7, 9], resonance Raman spectroscopy [10], points accumulation in nanoscale topography (PAINT) [11], Fourier ptychography [12], beam shaping and steering in free space [13] etc. For bio-imaging applications, wide waveguides (50-500 μm) are used and the field of view is limited by the imaging microscope objective. Such waveguides support multiple optical electromagnetic wave modes, where each mode represents an eigen solution of the wave propagation equation for the waveguide. These multiple modes can superimpose leading to multimode interference (MMI) in the waveguide core. This results in a non-homogenous evanescent field intensity distribution leading to uneven excitation of fluorescently labeled biological samples placed on top of the waveguide. However, by taking an average of several images, each image taken under a different combination of modes, a reduced modulation in intensity across the imaging area is obtained [14]. The different combination of modes can be generated by scanning the incident light spot on the input facet of the waveguide. However, the scanning process deteriorates the temporal resolution. Single mode waveguides can alleviate the problems caused by MMI, but due to the narrow width needed to excite the single mode, a long adiabatic taper would be needed to expand the mode for very wide waveguides (e.g. 500 μm). Using the adiabatic taper approach, the cross-section of the structure is gradually changed along the propagation direction of the light such that coupling of energy from the lower order mode into higher order modes is inhibited [15]. However, it poses challenges such as shadowing effects which are difficult to avoid [16]. These shadowing effects manifest as dark bands parallel to the direction of propagation of light and arise mainly due to strong localized scattering from the waveguide surface. The scattering could also arise due to material impurities or refractive index variations of the sample or in the waveguide itself. In this aspect, multiple mode illumination is advantageous as each mode illuminates any local region in the sample from different directions and reduces the shadowing effect.

Many studies nowadays revolve around the development and usage of intensity fluctuation-based algorithms which can improve spatial resolution over optical resolution limited fluorescence microscopy and temporal resolution over single molecule localizations methods. The spatio-temporal sparsity required by single molecule localization (SML) techniques [17, 18] for a reliable reconstruction demands large number of images and high laser power [5]. On the other hand, intensity fluctuation-based algorithms overcome these constraints but at the cost of relatively poorer spatial resolution than SML. Super-resolution optical fluctuation imaging (SOFI) [19], multiple signal classification algorithm (MUSICAL) [20], super-resolution radial fluctuations (SRRF) [21], entropy based super-resolution imaging (ESI) [22], sparsity based super-resolution correlation microscopy (SPARCOM) [23], Bayesian analysis of blinking and bleaching (3B) [24] can help generate super-resolved images using image stacks acquired from standard optical microscopes by using conventional fluorophores and nominal laser powers. These algorithms resort to higher order statistical analysis of intensity fluctuations from an emitter, a fluorophore molecule, as a function of time to generate super-resolved images.

In this article, we investigate the usage of one of the intensity fluctuation algorithms, namely SOFI, on waveguides to generate super-resolved images. In waveguide TIRF imaging, the intensity values recorded by a camera are the product of fluorophore distribution and the MMI pattern of the evanescent field. The fluctuations arise due to the intrinsic photokinetics of the fluorescent molecule and temporally varying non-uniform MMI pattern. These fluctuations which manifest as a change in intensity value at a particular pixel of a camera, can be localized to within subpixel precisions computationally using algorithms such as SOFI. However, it is observed that even though the average diffraction-limited image shows insignificant evidence of these MMI patterns, SOFI reconstructions provide prominent evidence of the non-uniform illumination, thereby hindering a reliable reconstruction. To alleviate this problem we investigate the usage of Haar wavelet kernel analysis (HAWK) [25] prior to applying SOFI. HAWK is a preprocessing algorithm that helps generate spatio-temporal sparse data sets via temporal band-pass filtering of the original data set. This helps in breaking the correlation in illumination pattern arising out of the MMI patterns. Here, a detailed experimental analysis is illustrated to generate chip based TIRF super-resolved images with minimized artifacts at high temporal resolution.

## 2. Methods

### 2.1 *Experimental setup*

In the chip-based microscope used in this paper, the core of the waveguide is made of a high refractive index (n) material, Tantalum pentoxide with n=2.1. Total internal reflection at the low refractive index boundary results into a 100-150 nm deep evanescent wave that exponentially decays away from the waveguide surface [7]. A CW laser (660nm Cobolt Flamenco, 561 nm Cobolt Jive) coupled to a single mode fiber is focused on the waveguide input face using a fiber-collimator and a microscope objective lens (Olympus LMPlanFL N, 50X/0.7 NA), allowing for end-facet coupling of light on the planar waveguide structure. The in-coupling optics are mounted on a piezo-electric XYZ translation stage. Using the high precision piezoelectric translation stage, the coupling optics can be shifted transversally along the waveguide input facet as shown by the blue arrows in Fig. 1(a), causing a spatial re-distribution of the guided modes. By shifting the MMI patterns in time, darker regions can also be illuminated and an average intensity distribution with reduced modulations across the entire waveguide area can be achieved. The fluorescently labeled sample is placed on top of the waveguide chip and excited via the evanescent field. The emitted Stoke shifted light from regions tagged by the fluorescent molecules is collected by an upright microscope fitted with emission filters and a sCMOS camera (Hamamatsu C11440-42U30). The experimental setup and waveguide TIRF concept are shown in Fig. 1(a) and 1(b).

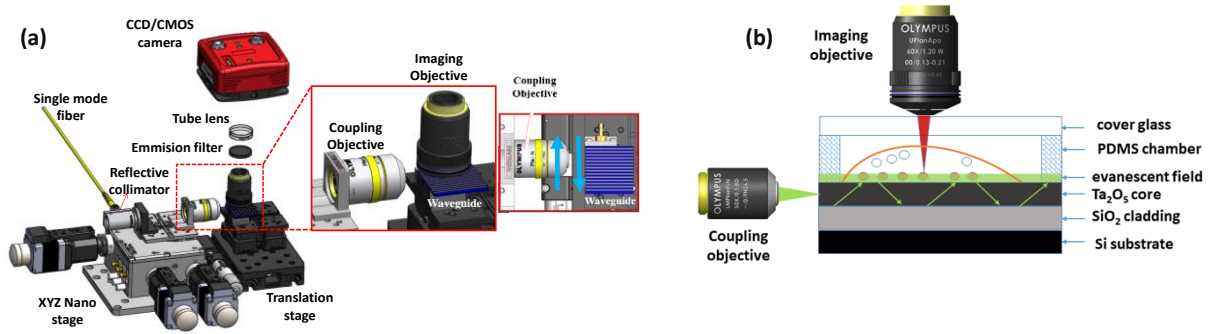

**Fig. 1.** Schematic diagram of chip-based imaging setup and waveguide TIRF concept. (a) Schematic diagram of waveguide based TIRF imaging is presented here. The blue arrows represent the oscillatory movement of the coupling objective along

the front facet of the waveguide. (b) A microscope objective (50X/0.7NA), referred to as coupling objective, is used to focus light into the core of a waveguide. The light is guided along the length of the waveguide via total internal reflection, shown using the green arrows. The fluorescently labeled sample to be imaged is placed on the top of the waveguide core. The evanescent field generated along the surface of the core as a consequence of total internal reflection illuminates a thin section of the sample in contact with the surface via the evanescent field. Only the fluorescent molecules (shown in red) in contact with the evanescent field are excited leaving the fluorophores outside the reach of evanescent field unexcited. The red shifted light emitted by the fluorescent molecules are collected by a collection objective. Waveguide TIRF approach dissociates the excitation and detection paths enabling scalable field-of-view.

## 2.2 Imaging configurations

The experimental results on chip-based intensity fluctuation imaging of tubulin in fixed PTk2 cells and actin in fixed merkel cell carcinoma cells (MCC13) are provided in this article. An image sequence of 300 frames for PTk2 cells and 500 frames for MCC13 cells is used as the input for the reconstruction algorithms. For PTk2 imaging, the waveguide is excited at 660 nm vacuum wavelength and images are acquired with an exposure time of 30 ms using an Olympus UPlanSApo 60X/1.2 NA water immersion objective referred to as the imaging objective. For MCC13 cells, the waveguide is excited using the 561 nm laser and images are acquired with an exposure time of 30 ms. The cells and aqueous imaging buffer are placed on top of the waveguide inside a polydimethylsiloxane (PDMS) chamber of ∼ 150 μm thickness. The chamber is sealed with a # 1.5 thickness coverslip and imaged using the imaging objective from top as shown in Fig. 1(b).

## 2.3 *SOFI and HAWK in the context of chip-based imaging*

Conventionally, the photokinetics of fluorescent molecules are exploited by fluctuation based algorithms for the generation of super-resolved images. Photokinetics in fluorescent molecules may arise due to blinking of the molecules as exploited in SML techniques or due to intrinsic spontaneous emission of molecules of the fluorescently labeled sample. As a consequence of photokinetics, the number of photons emitted by a fluorescent molecule in a given time duration is given by a probability density function, rather than a constant number, and manifests as changes in intensity values over time. The concept of photo kinetics in fluorescent molecules is depicted in Fig. 2.

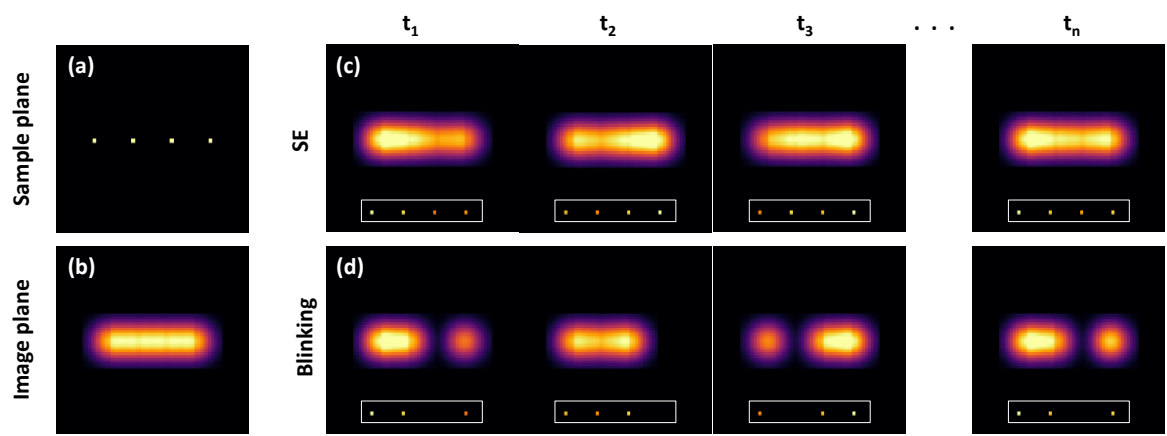

**Fig. 2.** Photokinetics panel. The different frames represents the images of four fluorophores recorded by the camera, generated due to convolution of the system PSF with the fluorophore shape in the sample plane. (a) Four emitters in the sample plane. (b) Image of the four emitters formed on the camera. The different frames acquired by the camera at different times $t_1$, $t_2$...$t_n$ are shown, where the time intervals are in the order of $10^{-3}$ s. The absorption of light by the fluorescent molecules takes place in the order of $10^{-12}$ to $10^{-15}$ s. The excited molecules may then relax to the ground state via radiative or a non-radiative transfer. The fluctuations necessary for generating super-resolved images may arise due to the following: (c) Spontaneous emission which is the intrinsic intermittent emission of the fluorophores, (d) blinking (on/off) of the fluorophores. The insets, not drawn to scale, in (c-d) show the corresponding intensity distribution of the four emitters in the sample plane.

A second order auto-correlation function may be expressed as shown in Eqn. 1 [19, 26]

$$G_2(\boldsymbol{r},\zeta) = \sum_{i,j=1}^{N} PSF(\boldsymbol{r}-\boldsymbol{r}_i).PSF(\boldsymbol{r}-\boldsymbol{r}_j).\varepsilon_i.\varepsilon_j.\langle \delta s_i(t+\zeta)\delta s_j(t)\rangle_t \quad (1)$$

where N is the number of emitters, PSF(**r**) is the point spread function of the system, $\varepsilon_{i,j}$ is the constant brightness of the i$^{th}$ and j$^{th}$ fluorescent molecules, $s_{i,j}(t)$ is the time-dependent fluctuation of these molecules, $\delta s_{i,j}$ quantifies the fluctuations over zero-mean and $\langle \cdots \rangle_t$ represents time averaging. If there is no correlation between the different emitters the above auto correlation function can be simplified to a 2$^{nd}$ order auto-cumulant function (n=2), see Eq. 2, and the pixel values in a SOFI image correspond to the cumulant values of the intensity distribution.

$$G_2(\boldsymbol{r},\zeta) = \sum_{i=1}^{N} PSF^2(\boldsymbol{r}-\boldsymbol{r}_i).\varepsilon_i^2.\langle \delta s_i(t)\delta s_i(t+\zeta)\rangle_t \quad (2)$$

Eq. 2 may be understood as pixel values in a SOFI image depending on the weighting terms $\varepsilon_i^2$ and $\langle \delta s_i(t)\delta s_i(t+\zeta)\rangle_t$. It implies that a SOFI image communicates about brightness and degree of correlation of temporal fluctuations in photon emissions from an emitter. A higher degree of fluctuation will yield a higher weighting factor and will be better visible in the SOFI reconstruction. This also implies that emitters with weaker weighting factors may get masked in the presence of brighter emitters. In the conventional application of SOFI, the resolution gain is achieved by assuming that a single emitter is spatio-temporally correlated with only itself. If the emission between the different emitters are mutually independent, then Eq. 2 can be invoked which leads to squaring of the PSF that will ensure super-resolution unlike the auto correlation function described by Eq. 1. Higher order cumulants may be employed to further enhance the resolution. A variation of SOFI, namely b-SOFI (balanced-SOFI), may be utilized to avoid the masking of weaker emitters, which linearizes the brightness to provide good contrast images without the weak emitters getting masked [22]. The SOFI reconstructions in this article are carried out using the MATLAB code, © 2012 Marcel Leutenegger et al, École Polytechnique Fédérale de Lausanne, under the GNU General Public License. In this article 2$^{nd}$, 3$^{rd}$, 4$^{th}$ order and b-SOFI reconstructions are performed for one dataset for comparing their performances. Otherwise, generally b-SOFI is used for other experiments. The different orders of the SOFI reconstructions correspond to the different orders of the auto-cumulants employed by the algorithm.

Using the waveguide platform for imaging, the fluorescent molecule fluctuations are controlled by the evanescent field distribution in each frame as shown in Fig.3. In each frame, the fluorescence is proportional to the excitation light intensity it receives as a consequence of the MMI pattern. The non-uniformity in illumination leads to correlation between the different fluorophores. This evanescent field illumination of fluorophores is analogous to random speckle illumination used in [28]. It implies that the assumption of no cross-correlation between different fluorophores as mentioned in [19] is violated. The non-uniform illumination induces correlation between the different fluorophores. This correlation between the emitters due to MMI patterns will modify the pixel values in the SOFI image according to Eq. 1. Therefore, the resolution enhancement achieved using SOFI on images acquired using a waveguide platform is analogous to resolution gain as in S-SOFI [28] after invoking contribution of the cross-correlation terms.

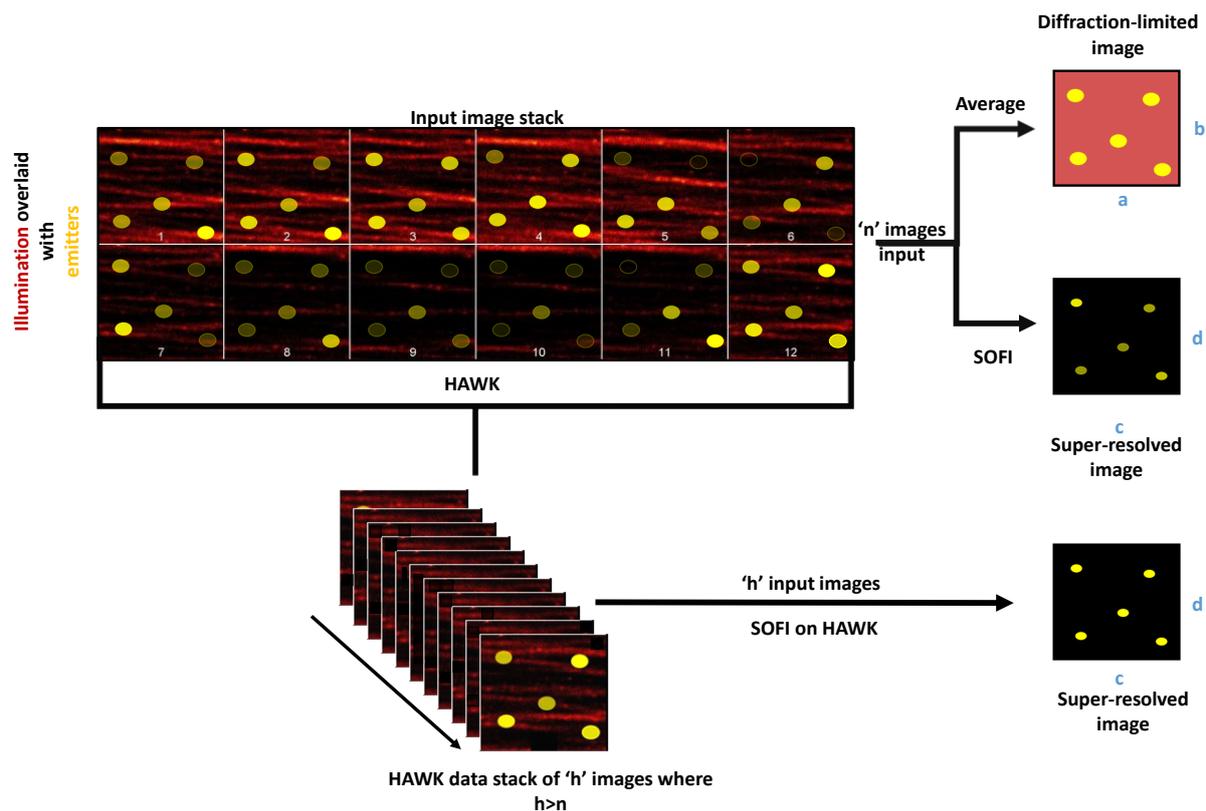

**Fig. 3.** Schematic representation of chip-based intensity fluctuation analysis. The fluctuations arising from the MMI pattern and the intrinsic fluctuation (SE) of the molecules are recorded in an image stack containing 'n' frames, with each frame of (a X b) pixels. The image stack of 'n' frames is duplicated and also preprocessed using HAWK. SOFI operates on both the original and HAWK data stack of diffraction-limited images to produce a super resolved image of (c X d) pixels where c > a and d > b. The non-uniform illumination leads to artifact generation in the super-resolved images generated using SOFI. These artifacts can be minimized by preprocessing the image stack with HAWK before applying SOFI.

HAWK (Haar wavelet kernel) analysis is a pre-processing algorithm that helps introduce computational sparsity into the original image stack via Haar wavelet transform (HWT). The intensity trace of a particular pixel over 'n' frames of the unprocessed image stack is expressed as a column vector A(t). Its transform intensity trace B(t') is synthesized as $B(t') = H.A(t)$, where H is the Haar matrix. Then a filter of level m is applied to B(t'), where m denotes the level of the Haar wavelet transform. This is achieved by setting to zero all the elements of B(t') that do not belong to that particular level m of the Haar matrix. The filtered pixel intensity trace is obtained via the inverse HWT and is given by $C^m(t) = H^T.B^m(t')$, where the inverse HWT is given by $H^{-1} = H^T$. Then a cropping procedure is applied as detailed in Ref. [25]. This process of transform-filter-inverse transform is performed on all the pixels of the image for the desired number of filter levels and the resulting image sequences are appended together to produce the Haar transformed data set. The application of filter levels and cropping procedure introduces zeros into the pixel intensity trace. In the case of chip-based imaging, this process helps to break correlation arising out of the non-uniform illumination. The breaking of correlation computationally leads to reduced pixel values in the SOFI image, thereby preventing masking of weaker emitters. The detailed theory describing the HAWK method can be found in [25].

## 3. Results and discussion

3.1 *HAWK helps break correlation and reduce average value of intensity trace per pixel*

An analogy between MMI in waveguides and speckle formation in free space optics is drawn. The high coherence of laser light gives rise to speckle phenomena and one of the methods of suppressing the speckle contrast is to image through a rotating diffuser [29]. The laser light is incident on a diffuser and light coming out of the diffuser is imaged on a camera. In this experiment, a diffuser is rotated sequentially from 1° to 360° in steps of 1° and an image of the speckle pattern is acquired after each 1° rotation. To illustrate the applicability of HAWK technique in reducing the overall average intensity value of the speckles, the original and HAWK data stacks are averaged in intensity as shown in Fig 4(b1-b2). Then the average value per column, i.e., along y-axis in Fig. 4, of the so obtained averaged images of size (432X432) pixels is calculated and is plotted in Fig. 4(c1) respectively. It can be seen from Fig. 4(c1) that through the introduction of HAWK the averaged image has a lower intensity value per pixel.

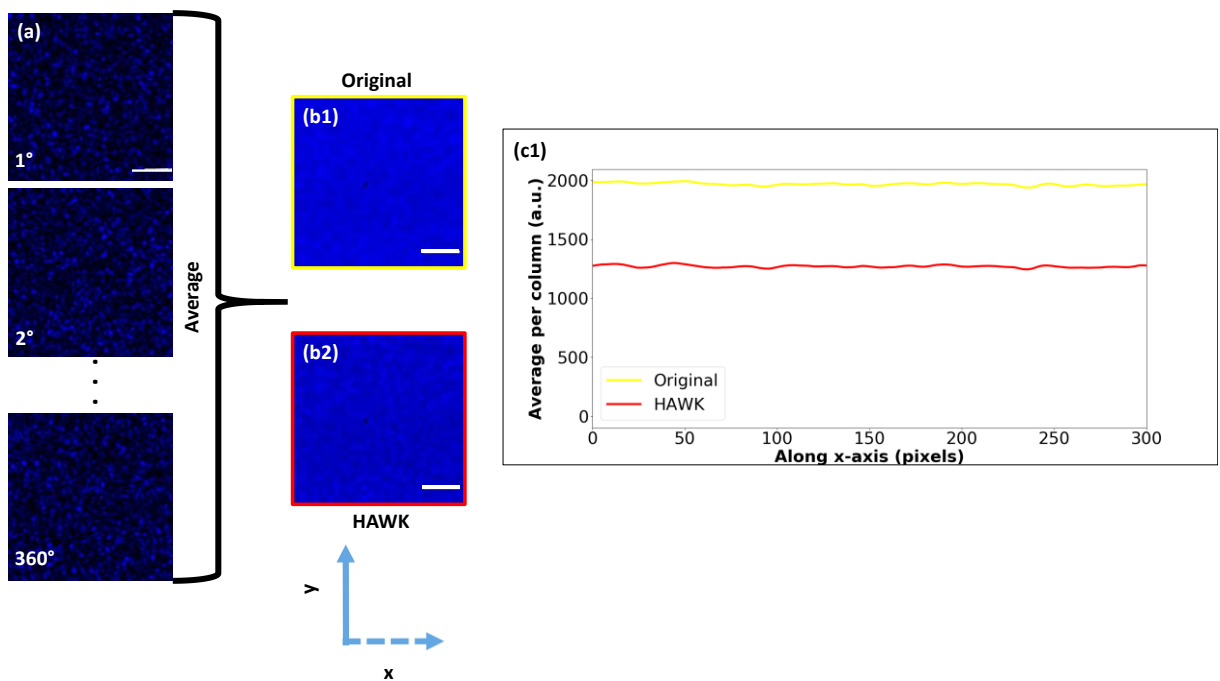

**Fig.4.** (a) The images of the speckles recorded after each 1° rotation of the diffuser is shown. A total of 360 images, of size 432 X 432, pixels are recorded corresponding to 360° rotation of the diffuser. (b1) Average intensity image of the original data stack and (b2) average intensity image of the HAWK data stack. (c1) Average value per column of the averaged original and HAWK data stacks. The dotted x-axis indicates the direction along the length of the waveguide and y-axis indicates the direction along the columns of the image. Scale bar 8 µm.

In waveguides, through the introduction of temporal sparsity HAWK helps to break the correlation arising out of MMI illumination. As can be seen from Fig. 4(c1) the average value per pixel is also reduced after the application of HAWK. This aids SOFI in reconstructing emitters exhibiting weaker fluctuations in the original data stack. However, this may reduce the signal-to-background ratio of the final reconstructed image. To study the influence of HAWK on breaking correlation in illumination, the MMI patterns of a waveguide are imaged. For this analysis, a 200 µm Tantalum pentoxide ($Ta_2O_5$) waveguide is coated with Alexa Flu647. A stack of 40 images is recorded. Each image is acquired with an exposure time of 30 ms as the coupling objective oscillates along the input facet of the waveguide while still maintaining coupling as described in Fig. 3. The image stack so acquired is duplicated into two. One stack is averaged in intensity using Fiji. The other stack is pre-processed using HAWK at level 3 and then averaged in intensity. Fig 5(a) represents an image stack of 40 frames acquired by oscillating the coupling objective. Fig. 5(b1-b2) represents average images of the original and HAWK data stack respectively. The average value and standard deviation of each column of the original and HAWK intensity-averaged images shown in Fig. 5(b1-b2) is calculated and shown in the plots Fig. 5(c1-c2) respectively. The yellow and red vertical arrows in Fig. 5(b1-b2) indicate the direction, i.e. along the columns of the image, in which average and standard deviation of the averaged image are calculated for original and HAWK data set respectively. The dotted arrows indicate the direction along the length of the waveguide. Similar to Fig. 4(c1), it can be seen that through the introduction of HAWK the averaged image has a lower intensity value per pixel for waveguide illumination as shown in Fig. 5(c1). The reduction in the standard deviation along the columns as depicted in Fig. 5(c2) signifies a more uniform illumination.

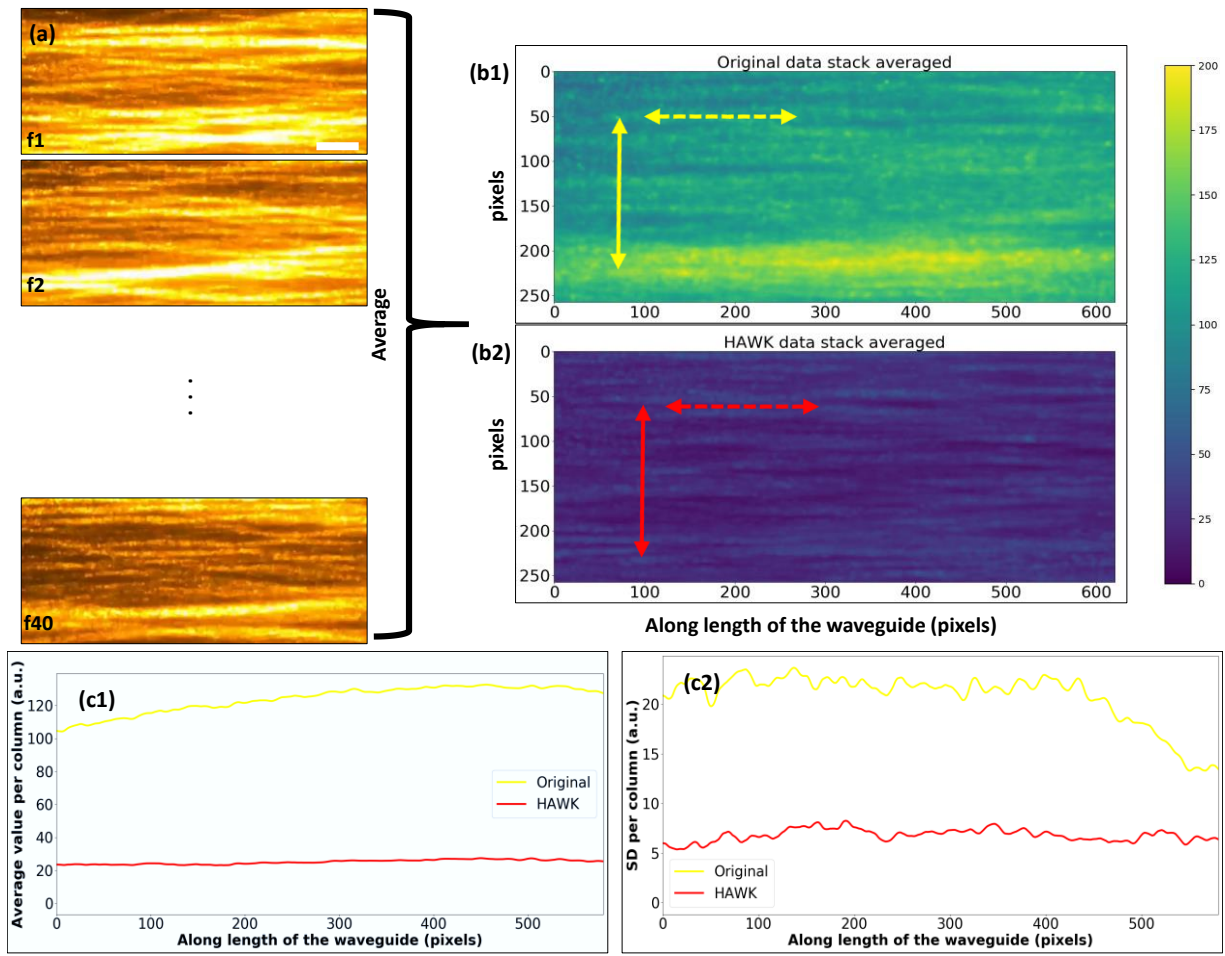

**Fig.5.** (a) Image stack of 40 frames acquired by oscillating the coupling objective. (b1) Average image of the original data stack, (b2) average image of the HAWK data stack. The vertical complete arrows indicate the direction of the columns and dotted horizontal arrows indicate the direction of rows, i.e., along the length of the waveguide. (c1) Average value and (c2) standard deviation of the columns of the original and HAWK averaged images. Scale bar 8 μm.

## 3.2 Chip based TIRF imaging

Tubulin filaments in PTk2 cells labeled using Alexa Flu647 is imaged in waveguide TIRF mode. The coupling objective, mounted on a piezo stage as shown in Fig. 1., is oscillated while sustaining coupling. A stack of 300 images is acquired at 30 ms per frame using a sCMOS camera.

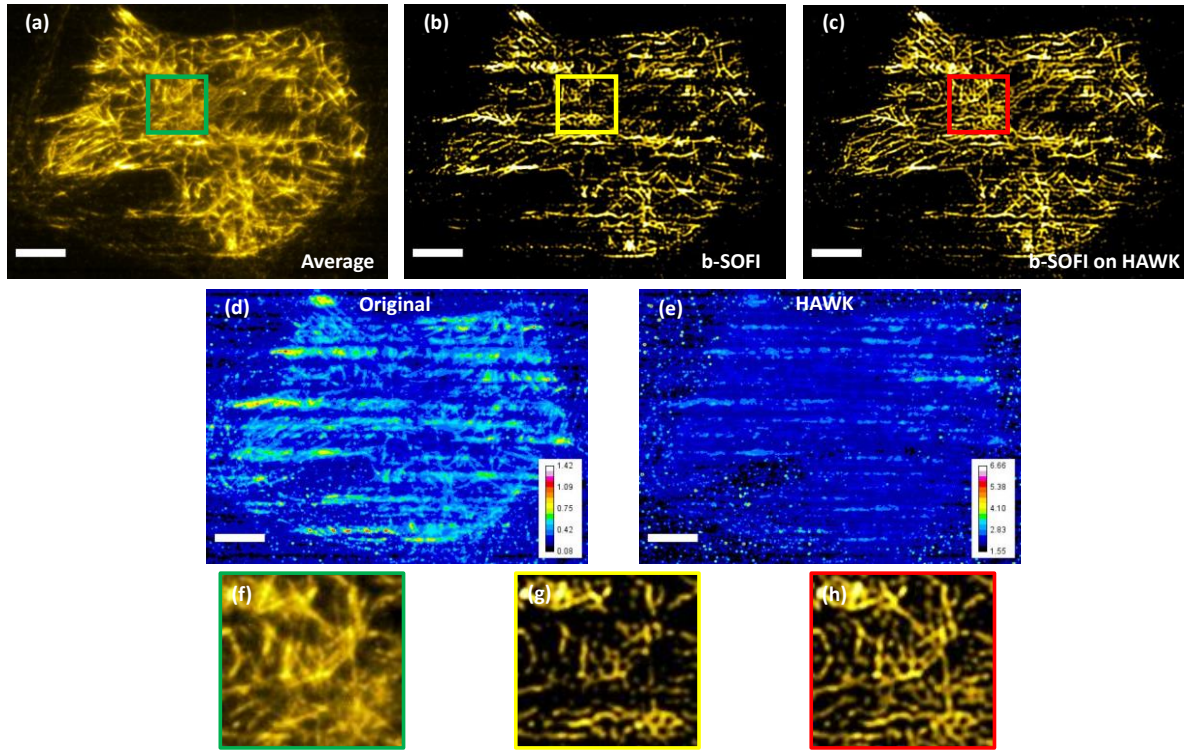

**Fig. 6.** Chip based TIRF images of tubulin in PTk2 cells. (a) Diffraction-limited TIRF image of tubulin in PTk2 cells generated by averaging an image stack of 300 frames. (b) b-SOFI reconstruction of original data stack (c) b-SOFI reconstruction of HAWK data stack. Ratio of standard deviation to average plot of (d) original data stack of 300 frames and (e) HAWK level 3 data stack of 1778 frames. The calibration bar quantifies the ratio of standard deviation to average taken for each pixel over all the frames. This ratio quantifies the strength of fluctuations in intensity over the average intensity. Blown up image of (f) green box shown in diffraction limited image, (g) yellow box shown in b-SOFI reconstruction of the original data stack and (h) red box shown in b-SOFI reconstruction of the HAWK data stack. Scale bar 8 μm.

An average diffraction-limited TIRF image is generated from the initial image stack of 300 frames. This is shown in Fig. 6(a). HAWK introduces artificial temporal sparsity and helps in depopulating densely packed regions, for example the green box shown in Fig. 6(a). The initial data stack of 300 frames is used to generate a HAWK level 3 data set of 1778 frames. Since SOFI relies on fluctuations for super-resolution, it is imperative to quantify the strength of the fluctuations. The standard deviation over the data stack at a particular pixel gives a measure of the strength of the fluctuations at that particular pixel. If a pixel hosts an emitter it shows fluctuations over time which are recorded in the image stack. Therefore, to quantify the strength of the fluctuations, the ratio of standard deviation to average for each pixel is computed over all the frames. This is done for both the initial data stack of 300 frames and HAWK data of 1778 frames and is shown in Fig. 6(d) and Fig. 6(e) respectively, and is referred to as fluctuation map in this article.

It can be seen from Fig. 6(b) that b-SOFI masks the weaker emitters due to the uneven illumination. But HAWK helps in preventing this masking of weaker emitters by breaking the correlation through the introduction of temporal sparsity. The effect is evident in b-SOFI reconstruction of the HAWK data set shown in Fig. 6(c). This can be understood from the fluctuation maps shown in Fig. 6(d-e). The fluctuation map of the original data stack shows the presence of very strong fluctuations from certain emitters. Even when b-SOFI is applied on such a data stack it leads to masking of the weaker emitters, i.e. emitters lying in the dark region of the MMI patterns over a longer course of time. However, the fluctuation map of the HAWK data stack reveals that by breaking the correlation between successive frames, HAWK decreases the average value. After the application of HAWK, the pixels will have a reduced value and an increase in the ratio of standard deviation to mean. An increase in standard deviation over mean means an increase in fluctuation in intensity in each pixel and from Eq.2. it is known that the pixel value of a SOFI image depends on fluctuations over mean. As a result, a pixel hosting an emitter will show more fluctuations and consequently more intensity in the SOFI reconstructed image than a pixel without an emitter. Therefore, even the weaker emitters can be picked up by b-SOFI after HAWK thereby leading to a more reliable reconstruction. The magnified images shown in Fig. 6(f-h) highlight its experimental verification.

The SOFI reconstructions on the original data set of 300 frames are given in Fig. 7(a-d) and the corresponding SOFI reconstructions on HAWK level 3 data set of 1778 frames is shown in Fig. 7(e-h). The resolution assessed using FRC is shown in Fig. 7(i-l). It is seen that SOFI reconstructions on HAWK data set yielded a better resolution and a more reliable reconstruction by preventing the masking of the weaker emitters. However, there is a tradeoff, as the signal to background ratio is lower for the HAWK data set.

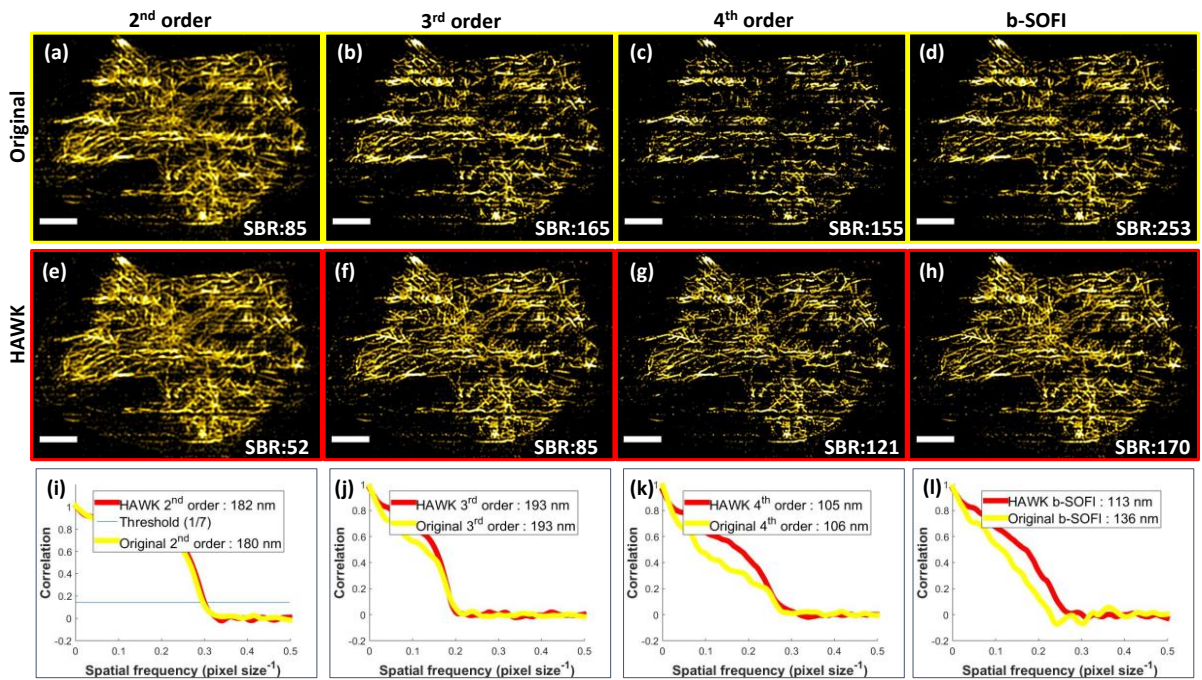

**Fig. 7.** Chip based SOFI imaging. SOFI reconstructions on original data set of 300 frames is shown by (a-d), enclosed within yellow frames. The orders of SOFI used for this purpose are (a) 2$^{nd}$ order, (b) 3$^{rd}$ order, (c) 4$^{th}$ order and (d) b-SOFI. SOFI reconstructions on HAWK data set of 1778 frames is shown by (e-h), enclosed within red frames. The SOFI orders used for reconstruction are (e) 2$^{nd}$ order, (f) 3$^{rd}$ order, (g) 4$^{th}$ order and (h) b-SOFI. Signal to background ratio (SBR) of the image is provided in the bottom right corner. The resolutions of the different reconstructions are quantified using FRC for (i) 2$^{nd}$, (j) 3$^{rd}$, (k) 4$^{th}$ and (l) b-SOFI. The FRC resolution values are provided as insets in the FRC plots. All the reconstructions are displayed here after scaling down using bilinear interpolation in Fiji. Scale bar 8 μm.

To showcase the strength of TIRF-imaging over large area using waveguide chip-based imaging platform for SOFI, MCC13 cells stained for actin are imaged using a 20X/0.45 NA and the region enclosed by the blue box in Fig. 8(a) is then imaged using a 60X/1.2 NA water immersion objective. In chip-based microscopy, the TIRF illumination is generated by the waveguide and is independent of the collection light path. Consequently, scalable field-of-view can be captured by using objective lens of different magnification. The results are shown in Fig. 8 and open up the possibility of generating super-resolved images using SOFI with minimized artifacts over large field-of-views.

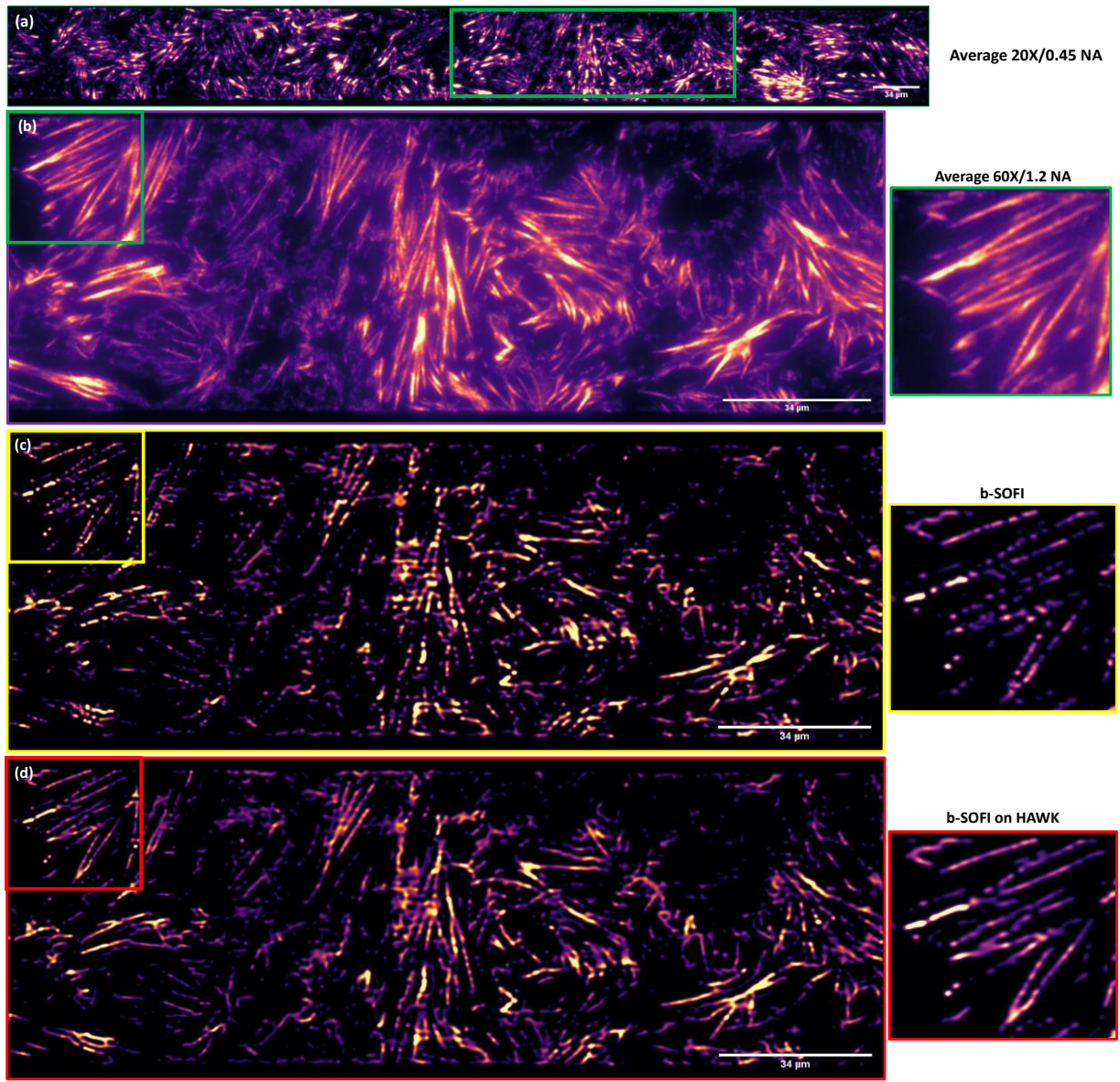

**Fig. 8.** Chip based SOFI imaging of MCC13 cells stained for actin (AF 555 phalloidin). (a) Diffraction limited TIRF image generated by averaging an image stack of 500 frames acquired using 20X/0.45 NA. The region shown inside the green box is (b) imaged using a 60X/1.2NA water immersion objective and shown enclosed in a purple frame, (c) reconstructed using b-SOFI on the original data set of 500 frames and shown enclosed in a yellow frame with FRC resolution of 299 nm, and (d) reconstructed using b-SOFI on the HAWK data set of 2978 frames and shown enclosed in a red box with FRC resolution of 241 nm. Scale bar 34 µm.

## 4. Conclusion

The employment of SOFI on chip-based imaging platforms helps gain resolution over a large field-of-view. It was observed that many of reconstruction artifacts are not due to SOFI, rather due to the on-chip illumination scheme employed. Wide waveguide generates an uneven distribution of the evanescent field arising due to multiple modes propagating simultaneously in the waveguide. To overcome this challenge, application of HAWK on the image stack prior to the application of SOFI has proven to be useful. In particular, the challenges associated with the masking of the weaker emitters showed improvements. Though the artifacts are not completely eliminated they have been minimized. Future work will focus on designing multi-mode waveguide structures with Y-junctions as shown in Ref. 30 to generate an almost uniform evanescent field illumination. We are also investigating the fact that a better labeling strategy of the cells, i.e. by minimizing the bleeding of dyes into the background, might help in improving the SBR lost after the application of HAWK. Our preliminary results suggest that the concept of application of HAWK to minimize the artifacts arising due to MMI patterns in waveguides can also be extended to other algorithms like MUSICAL, super-resolution method based on auto-correlation two-step deconvolution (SACD) [31] etc.

As compared to chip-based SMLM approaches [7, 11], chip-based SOFI is an effective technique to generate super-resolved images with relatively higher temporal resolution. Waveguide platform has been previously used TIRF microscopy on living cells [14] and super-resolution imaging of fixed cell using SMLM method [9]. Future work will focus on super-resolution imaging of living cells exploiting chip-based SOFI. As chip-based SOFI needs similar number of images, a few hundred images, as acquired in chip-based TIRF, the method is suitable for live cell imaging application. Furthermore, chip-based SOFI can easily be integrated with other on-chip optical functions such as on-chip Raman spectroscopy [32-34], waveguide trapping [35-38], optical phase tomography [39-41] and others.

**Funding**

H2020 Marie Skłodowska-Curie Actions (MSCA_ITN: 31147)

KA acknowledges funding from a Horizon 2020 Marie Skłodowska-Curie Action (SEP-210382872) and a Horizon 2020 ERC Starting Grant (804233).

**Acknowledgements**

The authors would like to thank Dr. Deanna Wolfson, Dr. Firehun T Dullo, Dr. Azeem Ahmad and Dr. Vishesh Dubey for their valuable inputs during the course of this work.

**Disclosures**

B.S.A. have applied for patent GB1606268.9 for chip-based optical nanoscopy. B.S.A and O.I.H. are co-founders of the company Chip NanoImaging AS, which commercializes on-chip super-resolution microscopy systems.